\documentclass{aa} 
\usepackage{graphicx}
\usepackage{txfonts}
\usepackage{natbib}

\bibpunct[; ]{(}{)}{;}{a}{}{}

\begin{document}

   \title{A search for p-modes and other variability in the binary system 85\,Pegasi using 
   MOST\thanks{Based on data from the MOST satellite, a Canadian Space Agency mission,
jointly operated by Dynacon Inc., the University of Toronto Institute for Aerospace Studies and the 
University of British Columbia, with the assistance of the University of Vienna.} photometry}

	\titlerunning{MOST photometry of 85\,Pegasi}
	\authorrunning{D. Huber et al.}

   \author{D. Huber\inst{1,10}
		\and J.M. Matthews\inst{2}
		\and B. Croll\inst{3}
		\and M. Obbrugger\inst{1}
		\and M. Gruberbauer\inst{1}
		\and D.B. Guenther\inst{4}
		\and W.W. Weiss\inst{1}
		\and J.F. Rowe \inst{5}
		\and T. Kallinger\inst{1}
		\and R. Kuschnig\inst{1}
		\and A.L. Scholtz\inst{6}
		\and A.F.J. Moffat\inst{7}
		\and S. Rucinski\inst{3}
		\and D. Sasselov\inst{8} 
		\and G.A.H. Walker\inst{9}}
		
	\offprints{Daniel Huber, huber@astro.univie.ac.at}

   \institute{Institute for Astronomy, University of Vienna,
              T\"urkenschanzstrasse 17, 1180 Vienna, Austria
		\and Dept. of Physics and Astronomy, University of British Columbia, 6224 Agricultural Road, Vancouver, BC V6T 1Z1, Canada 
		\and Dept. of Astronomy and Astrophysics, University of Toronto, 50 St. George Street, Toronto, ON M5S 3H4, Canada
		\and Dept. of Astronomy and Physics, St. Mary's University, Halifax, NS B3H 3C3, Canada 
		\and NASA-Ames Research Park, Mail Stop 244-30, Moffett Field, CA 94035-1000, USA
		\and Institute for Communications and Radio-Frequency Engineering, Vienna University of Technology, 
		Gusshausstrasse 25-29, Vienna, Austria
		\and D\'ept. de physique, Univ. de Montr\'eal C.P. 6128, Succ. Centre-Ville, Montr\'eal, QC H3C 3J7, Canada  
		\and Harvard-Smithsonian Center for Astrophysics, 60 Garden Street, Cambridge, MA 02138, USA 
		\and 1234 Hewlett Place, Victoria, BC V8S 4P7, Canada
		\and Sydney Institute for Astronomy, School of Physics, University of Sydney, NSW 2006, Australia
			}
             
   \date{Received; accepted}

  \abstract
  % context heading (optional)
  % {} leave it empty if necessary  
   {Asteroseismology has great potential for the study of metal-poor stars due to its sensitivity to 
   determine stellar ages. Solid detections of oscillation frequencies in stars with well constrained 
   fundamental parameters, combined with a known rotation period, should 
   significantly advance our understanding of stellar structure and evolution in context with metallicity effects.}
  % aims heading (mandatory)
   {Our goal was to detect p-mode oscillations in the metal-poor sub-dwarf 
   \makebox{85\,Peg\,A} and to search for other variability on longer timescales.}
  % methods heading (mandatory)
   {We have obtained continuous high-precision optical photometry of the binary system 85\,Pegasi
   with the MOST (Microvariability \& Oscillations of STars) space telescope in two seasons (2005 \& 2007).
   The light curves were analyzed using traditional Fourier techniques. Furthermore, we redetermined 
   $v\sin i$ for 85\,Peg\,A using high resolution spectra obtained through the ESO archive, and 
   used photometric spot modeling to interpret long periodic variations.}
  % results heading (mandatory)
   {Our frequency analysis yields no convincing evidence for p-modes significantly above a noise level of
   4\,ppm. Using simulated p-mode patterns we provide upper RMS amplitude limits for \makebox{85\,Peg\,A}. 
   After removal of instrumental trends 
   the light curve shows evidence for variability with a period of about 11\,d and this periodicity is 
   also seen in the follow up run in 2007; however, as different methods to remove instrumental trends 
   in the 2005 run yield vastly different results, the exact shape and periodicity of the 2005 
   variability remain uncertain. Our 
   re-determined $v\sin i$ value for 85\,Peg\,A is comparable to previous studies and we provide 
   realistic uncertainties for this parameter. Using these values in combination 
   with simple photometric spot models we are able to reconstruct the observed variations.} 
   % conclusions heading 
   {The null-detection of p-modes in \makebox{85\,Peg\,A} is consistent with theoretical values 
   for pulsation amplitudes in this star. The detected long-periodic variation in the 85\,Peg system 
   must await confirmation by further observations with similar or better precision and long-term 
   stability. If the 11\,d periodicity is real, rotational modulation of surface features on one of the components 
   is the most likely explanation.} 
  
   \keywords{techniques: photometric -- stars: individual: HD\,224930 -- stars: individual: 85\,Peg -- 
   stars: oscillations -- stars: rotation -- starspots -- binaries:general}

   \maketitle

\section{Introduction}

With observations dating back to the mid-19$^{\rm th}$ century by \citet{baily}, 85\,Pegasi (HD\,224930) is a very well studied 
visual binary system with an angular separation of 0.83 arcseconds.  Due to its proximity ($d=12$\,pc) and 
brightness ($V=5.75$), the orbital elements have been determined to high accuracy through 
photometry and spectroscopy.  \citet{griffin} provides a comprehensive review of the 
wealth of data obtained for the \makebox{85\,Peg} system, including a spectroscopic orbital solution 
with an orbital period of 26.31 years and a total mass of $1.49\,M_{\sun}$. The dynamically 
measured relative masses of the binary system, however, indicating a secondary equally or only slightly 
less massive than \makebox{85\,Peg\,A}, disagree with the observed magnitude difference of 
${\Delta}m_{\mathrm{V}}\simeq 3$ 
between the components. It was suggested as early as 1948 by \citet{hall} that the secondary star in 
\makebox{85\,Peg} may be itself a binary system.  Evolutionary models by \citet{fernandes}, 
\citet{dantona} and \citet{bach} all support the binarity of \makebox{85\,Peg\,B}.

The primary \makebox{85\,Peg\,A} is metal-poor, with measurements of $\mathrm{[Fe/H]}=-0.88$ \citep{allende}, 
$-0.78$ \citep{holmberg} and  $\mathrm{[M/H]}=-0.69$ \citep[van't Veer 2000; cf.][]{fernandes}. The space velocity of the star, 
along with its low metallicity, qualify it as a sub-dwarf. 
\makebox{85\,Peg\,A} was classified as a G5 sub-dwarf by \citet{fulbright}.  Based on only the visual 
magnitude difference (neglecting the contradicting mass ratio from the orbital solution), \makebox{85\,Peg\,B} 
is usually assumed to be a K6--8 dwarf. 

\citet{fernandes} modeled the initial abundances, age and mixing-length parameters of 
\makebox{85\,Peg\,A} and estimated an age of $9.3\pm0.5$\,Gyr. Using effective temperatures derived from 
spectroscopy and the measured Hipparcos parallax, \citet{dantona} computed evolutionary 
tracks for a set of possible metallicities and ages of the system. Their evolutionary models allow 
ages between about 8 and 14\,Gyr. They generated pulsational eigenspectra from their models, predicting 
that solar-like p-modes of low degree ($\ell=0$--3) in \makebox{85\,Peg\,A} could occur across a frequency 
range of 1--5\,mHz ($\simeq 80$--$400$\,d$^{-1}$). Table \ref{tab:paras} lists all important parameters 
concerning the binary system 85\,Peg which are relevant for this paper. 
%A complete listing of all important orbital and physical parameters can 
%also be found in Tables 1 and 2 of \citet{dantona}.

\begin{table} 
\caption{Main parameters of 85\,Peg\,A and B compiled from the literature. Values in brackets for 
component B are suggested values by \citet{dantona} for the suspected binary system 85\,Peg\,Ba and Bb. 
Reference coding: 
(1) \citet{tenbrum}, (2) \citet{fulbright}, (3) \citet{griffin}, (4) \citet{dantona}, (5) \citet{bach}, 
(6) \citet{fernandes}, (7) van't Veer et al. (2005); cf. \citet{dantona}, (8) \citet{holmberg}}
\centering                      
\begin{tabular}{c c c c}        
\hline         
\hline
parameter					&	85\,Peg\,A		&		85\,Peg\,B [a,b] 						& Ref		\\ 
\hline
$V$ [mag]			&		5.81\,$\pm$\,0.03		&		8.89\,$\pm$\,0.29					&	1		\\
Sp.Type				&		G5\,IV					&		K6--8\,V								&	2		\\
$M/M_{\sun}$		&		0.77\,$\pm$\,0.05		&		0.72\,$\pm$\,0.05 [0.5,0.2]			&	3,4,5	\\
$L/L_{\sun}$		&		0.617\,$\pm$\,0.02		&		0.072\,$\pm$\,0.03 [0.064,0.0085]	&	4,6		\\
$T_{\rm eff}$ [K]	&		5600\,$\pm$\,50			&		4200\,$\pm$\,200					&	6		\\
$\log g$			&		4.6\,$\pm$\,0.1			&		4.8\,$\pm$\,0.2						&	4,7		\\
 $\mathrm{[Fe/H]}$	&		$-0.78$					&		-									&	8		\\
age [Gyr]			&		9.3\,$\pm$\,0.5			&		-									&	4		\\
\hline
\end{tabular} 
\label{tab:paras} 
\end{table}

The detection of oscillations in stars like 85\,Peg\,A is particularly interesting due 
to the sensitivity of asteroseismology to determine stellar ages. While the small frequency separation 
is the more sensitive observable for this purpose, even a measurement of the large separation considerably constrains 
the parameter space which often is more uncertain for metal-poor objects than for 
other stars.  The first successful application of asteroseismology to constrain the age 
of a metal-poor star has been performed by \citet{bedding_nuindi} using Doppler velocity observations of 
the subgiant  $\nu$\,Indi ($\mathrm{[Fe/H]}=-1.4$). Even before that, 
\citet{mosser_2005} detected oscillations in velocity measurements of HD\,49933 ($\mathrm{[Fe/H]}=-0.4$) 
which were confirmed photometrically by the CoRoT space telescope \citep{app_hd49933,michel_science}. 
The rich oscillation spectrum of HD\,49933 has recently been subject to some discussion, 
in particular in context with its fundamental parameters \citep[Bruntt et al. 2009, accepted]{bruntt2008,ryab} and the ability to model 
the oscillation frequencies \citep{kall_hd49933}, underlining the importance of studying stellar oscillations in 
metal-poor objects with well constrained parameters. Further 
recent detections of solar-like oscillations in metal-poor stars include the CoRoT timeseries of 
HD\,181906 ($\mathrm{[Fe/H]}=-0.1$) \citep{michel_science} and Doppler measurements of HD\,203608 
\citep{mosser_2008} ($\mathrm{[Fe/H]}=-0.65$).

The detection of p-modes and the measurement of the large and small frequency spacings in the 
high-overtone eigenspectrum  of \makebox{85\,Peg\,A} would put important constraints on its mass and age. 
If the star were as old as 12\,Gyr or more, this would also place a meaningful lower limit on the 
age of the Galaxy. For these reasons, even before the work of \citet{fernandes} and 
\citet{dantona}, \makebox{85\,Peg} was selected as a primary science target for the MOST space 
mission, to search for p-modes in the metal-poor sub-dwarf component of the binary. The continuous 
time coverage, high sampling rate and high photometric precision of the MOST photometry also make 
the data suited to search for other variability in the binary system.

\section{MOST observations \& data reduction}

The MOST (Microvariability \& Oscillations of STars) space mission is the first satellite designed 
to specifically conduct asteroseismic observations of stars \citep{walker03,matthews07}. 
Equipped with a 15\,cm Rumak-Maksutov telescope feeding a CCD photometer through a custom broadband 
optical filter (350--700\,nm), MOST can collect rapid high-precision photometry of bright stars 
with nearly uninterrupted coverage for up to 2 months. 

Depending on the brightness of the target of interest, MOST uses two different observing modes. 
In the primary mode, the target starlight enters a field stop about 1\,arcmin in diameter and 
illuminates an extended fixed image of the telescope pupil on the science CCD detector using a Fabry 
microlens. For fainter stars, direct photometry of stars is obtained corresponding to traditional 
CCD photometry with a FWHM of about 2--3 pixels. With a pixel scale of $\sim 3$ arcsec, the binary 
nature of 85\,Peg remains unresolved (the integrated signal, however, being 
dominated by \makebox{85\,Peg\,A} which is about $16\times$ brighter than its companion based on the observed 
visual magnitude difference).

MOST observed 85\,Peg in 2005 and 2007. In 2005, the visual brightness limit for the primary 
observing mode was $V\sim 6$, making 85\,Peg ($V=5.75$) the faintest star ever observed using Fabry imaging. 
In early 2006, the satellite tracking CCD stopped functioning, most likely due to a strong particle hit. 
Consequently, guide star observations had to be relocated to the science CCD, limiting the exposure times 
for the program stars and constraining the brightness limit for Fabry Imaging. Therefore, 85\,Peg was observed 
using direct imaging in the follow-up run in 2007. 

For both observation modes, data reduction pipelines have been developed and are fully described 
by \citet{reegen06} for Fabry imaging and \citet{rowe} as well as \citet{huber} for direct imaging. 
The main issue in MOST data reduction is the removal of stray light influences from the data. In all 
programs, the concept of decorrelation is used which is based on the comparison of intensities 
originating from the star with intensities of background readings. The 85\,Peg datasets were 
reduced according to the routines described by \citet{reegen06} and \citet{huber}.
%In the routines of \citet{reegen06} and 
%\citet{huber}, which were used in this analysis, a correlation is calculated over a time 
%subset of typically 5 MOST orbits ($\sim$ 0.3\,d) and corrected by calculating and subtracting a linear 
%or polynomial fit to this correlation. Decorrelation is performed for mean intensities, but also as a stepwise routine 
%with single pixel intensities, using the background pixel showing the best correlation coefficient 
%with the mean target pixel values. While the stability of Fabry imaging to small satellite jitter effects 
%allows a simple aperture definition, the direct imaging reduction software implements an additional 
%shifting routine to align all stellar PSFs for a fixed definition of pixels containing stellar intensities 
%and background readings, respectively. In addition to the stray light correction, both programs implement 
%outlier rejections based on the evaluation of the intensity distribution on the CCD, as well as an
%identification and correction of cosmic ray events. 

85\,Peg was observed by MOST in 2005 for 25.6 days with almost no interruptions. 
The exposure time was 50 seconds with a sampling interval of 55 seconds. After outlier 
corrections in the reduction pipeline the final light curve consisted of 34,323 measurements in 
total. The reduced light curve shows very good quality with point-to-point scatter of 0.48\,mmag, 
making it well suited to search for p-mode oscillations. In 2007, MOST returned to 85\,Peg for another 
run spanning a total of 25.5 days, this time with observations being performed in direct imaging mode with 
considerably shorter exposure times and images mostly being obtained every fifth MOST orbit. Due to the more 
difficult thermal conditions caused by target switching, the data are influenced by strong instrumental 
trends, which in particular affected the beginning and end of the run. Therefore, only the central part of 
the light curve spanning 
over about 12 days was used for the analysis. The data also show a slightly higher scatter than the earlier run, 
with an increased point-to-point spread of the light curve binned to about the same exposure time as 
the 2005 data. An observation log for both runs is given in Table \ref{tab:obslog}.

\begin{table} 
\caption{Observation log of the MOST 85\,Peg runs in 2005 and 2007. Values in brackets 
for the 2007 run correspond to the light curve with 4 datapoint bins (in order to make 
the scatter comparable to the 2005 run).} 
\centering                      
\begin{tabular}{l c c}        
\hline         
\hline
observation run					&	2005			&		2007 		\\ %					2007			\\
\hline
start date [HJD-2451545]		&	2085.43			&		2830.81 	\\ %				2826.52			\\
end date [HJD-2451545]			&	2111.01			&		2842.50		\\ %				2851.97			\\
total obs. time [d]				&	25.58			&		11.69		\\ %				25.45			\\
number of exposures				&	34,323			&		5\,738		\\ %				24,581			\\
duty cycle [\%]					&	85		&		14			\\ %				28				\\
exposure time [sec]				&	50				&		13.25 [53]	\\
sampling time [sec]				&	55				&		25 [100]	\\
point-to-point scatter [mmag]	&	0.48	&		0.83 [0.51]	\\		
\hline
\end{tabular} 
\label{tab:obslog} 
\end{table}

\section{The search for p-modes}

\subsection{Frequency analysis}

Due to the data quality difference, the search for p-modes was only performed in the 2005 data, using 
Fourier analysis and least squares fitting. 
The significances of identified peaks were estimated with {\sc{SigSpec}} \citep{reegen07} which, based on a 
false-alarm probability, incorporates frequency, amplitude and phase information into the 
calculation. Significant frequencies are identified through consecutive prewhitening and least 
squares fits. The analysis was performed up to the Nyquist frequency of the data set (${\nu}_{\rm 
Nyq}\simeq 9$\,mHz\,$\simeq 780$\,d$^{-1}$).

\begin{figure}
\resizebox{\hsize}{!}{\includegraphics{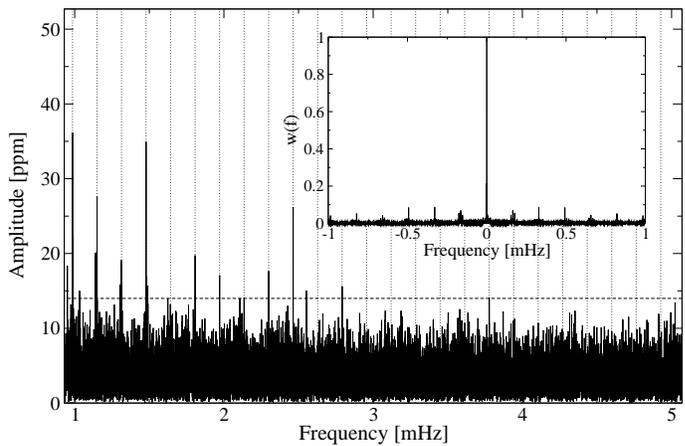}} 
\caption{Fourier amplitude spectrum of the 2005 MOST light curve of 85\,Peg in the frequency region where 
p-modes have been predicted. Vertical dotted lines correspond to harmonics of the 
orbital frequency of the MOST satellite. The horizontal dashed line shows the 3.5\,$\sigma$ detection 
limit (corresponding roughly to a spectral significance of 4) for the noise level in this frequency region. 
The insert displays the window function of the light curve.}
\label{fig:fig01}
\end{figure}

Figure \ref{fig:fig01} shows the Fourier amplitude spectrum in the frequency region where 
p-modes have been predicted. The amplitudes have been normalized by $N/2$ where $N$ is the 
number of datapoints in the light curve. The largest peaks in the amplitude spectrum are residuals of the background 
variation at the orbital frequency of the MOST satellite (164\,$\mu$Hz = 14.19\,d$^{-1}$) and 
its harmonics, marked in Figure \ref{fig:fig01} by vertically dotted lines. Note that even the largest of 
these has an amplitude of only about 35\,ppm (0.035\,mmag), and above a frequency of 2.5\,mHz, there 
is only one with an amplitude as high as 15\,ppm. There is also some power in sidelobes of the harmonics 
of the satellite orbital frequency, spaced by 1 and 2\,d$^{-1}$, due to the daily modulation of scattered 
earthshine.

Aside from these instrumental frequencies, there is no apparent power excess above the noise level 
visible in the spectrum. The {\sc{SigSpec}} analysis shows that, above 0.57\,mHz (50\,d$^{-1}$), there 
are only two peaks (not associated with MOST orbital harmonics or 1\,d$^{-1}$ sidelobes) with a 
significance greater than 4 (corresponding to an amplitude S/N of about 3.5). One of these peaks can be seen in 
Figure \ref{fig:fig01} near 2.55\,mHz.

\subsection{Estimation of the detection limit}

The MOST observations constitute the first large-scale dedicated observing effort to detect p-mode oscillations 
in the 85\,Peg system. Despite the null-result, the data should allow us to draw valuable 
conclusions for future efforts by setting firm limits on the possible pulsation properties of 
85\,Peg\,A.

Considering the average noise level of 4\,ppm in the amplitude spectra at a frequency range where the 
p-modes are predicted by theory, we can set an upper limit of pulsation in the MOST observations in 
Fourier domain to about 14--18\,ppm, assuming a 3.5--4.5\,$\sigma$ detection limit. Simulations of 
11 coherent sinusoidal variations inserted into the MOST photometry spaced by ${\Delta}\nu=170$\,$\mu$Hz (a 
value lying in the range of the models calculated by \citet{dantona}) and centered around 3.4\,mHz (using 
the relation of \citet{kjeldsen} for the estimated mass, radius and effective temperature of 
\makebox{85\,Peg\,A}) showed that peaks with amplitudes of 15\,ppm are securely recovered using the 
same approach as in the real data using {\sc{SigSpec}}. Note that this limit is only valid 
under the assumption that the mode lifetimes are longer than the observation timebase and therefore 
can be modelled as coherent sinusoidal variations.

Going one step further, we have also included the nature of the pulsation in the simulations. It is 
widely accepted that p-mode oscillations in sun-like stars are driven by turbulent motions in the 
convective layers in the stars, causing a stochastic excitation and subsequent damping of the 
pulsation modes rather than a coherent oscillation. It has been shown that 
the lifetimes of the oscillations for sun-like stars are very short, i.e. on the order of a couple of days 
\citep[e.g. for $\alpha$\,Cen\,A, see][]{bedding}, while both short as well as substantially 
longer lifetimes (i.e. on the order of weeks or months) have been detected in giant and subgiant 
stars \citep{stello,carrier,kallinger}.
While the situation of subdwarf stars remains undetermined, it seems appropriate from these results that with a 
dataset as long as the MOST observations the effects of stochastically driven pulsation should not be neglected.

The main effect when analyzing such signals in the Fourier domain is that the oscillation will not be 
described purely by the window function of the dataset, but rather by a series of peaks scattered 
around the true frequency value. The shape of this distribution is commonly described by a Lorentzian 
profile with the FWHM of the profile being correlated to the mode lifetime and the 
height, in combination with the width, being related to the total energy of the pulsation mode 
\citep{houdek}. Hence, the resulting amplitude of a mode in a Fourier spectrum is not only simply dependent 
on the maximum displacement in the time series, but also on the amount of damping and the resulting 
mean variation in the data set. As given by \citet{baudin} and \citet{barban}, the mean mode 
amplitude $A$ can be expressed as

\begin{equation}
A^{2} = \frac{T}{4 \tau} H = \frac{1}{4\tau} \mathcal{H}
\label{equ:amp}
\end{equation}
\noindent
with $T$ denoting the effective length of the dataset, $\tau$ is the mode lifetime, $H$ the height of the 
Lorentzian profile in power (ppm$^{2}$) and $\mathcal{H}$ the height of the Lorentzian profile 
in power density (ppm$^{2}$\,Hz$^{-1}$)\footnote{Note that this relation is only valid when the amplitude spectrum is normalized 
by $N/2$, where $N$ is the number of datapoints in the time series.}. The effective dataset length $T$ 
is evaluated as the reciprocal value of the area under the spectral window (in power), and in the 
case of the MOST 85\,Peg observations gives $T=1.8895$\,Msec $=21.869$\,d.

\begin{figure}
\resizebox{\hsize}{!}{\includegraphics{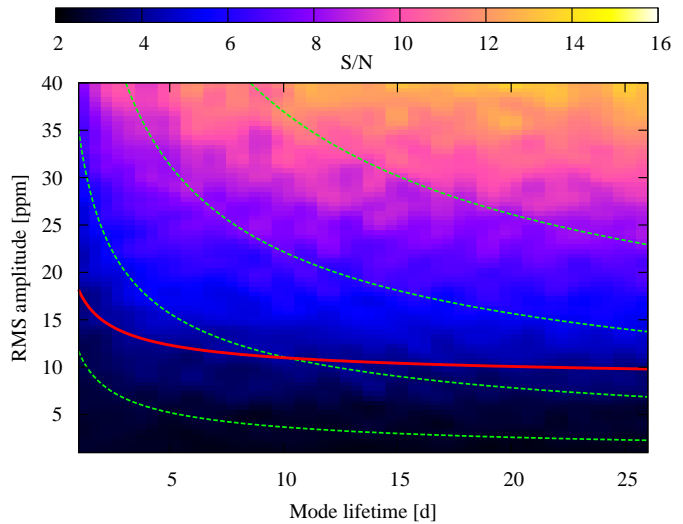}} 
\caption{Signal-to-Noise ratio of the six highest peaks of simulated solar-like pulsation using 
the MOST 85\,Peg data as a function of mode lifetime and RMS amplitude in the time domain. 
Points of constant Lorentzian profile height $H$ are shown as dashed green lines (from the top right to 
the bottom left: 50\,ppm, 30\,ppm, 15\,ppm and 5\,ppm). The thick solid red line shows a power law 
fit to bins of data points with S/N=3.5--4.5.}
\label{fig:fig02}
\end{figure}

To investigate limits on the mean mode amplitude for 85\,Peg\,A, we simulated solar-like oscillations 
implementing the concept by \citet{chaplin} adapted to the sampling of the MOST 85\,Peg 2005 time series. 
Using the same input frequencies as in the simulations containing coherent signal as described above, about 
1000 datasets were generated for mode lifetimes ranging from 1--26\,d in steps of 1\,d and for mean 
amplitudes from 1--40\,ppm in steps of 1\,ppm. Note that in order to keep the computation time reasonable and to 
solely study the effects of different amplitudes and mode lifetimes, each simulation was computed 
as a single realization (i.e. the stochastic excitation/damping pattern was the same for each 
simulated data set). 
The mean mode amplitude (which was set to the same value for all 
frequencies) was simulated by setting the time-domain standard deviation of a single realization to a fixed value 
$A_{\rm rms}$. Note that this value is not the same as the maximum peak height in the resulting Fourier 
spectrum, which is a priori unknown. In the following, the standard deviation of the variation in 
the time domain will be denoted RMS amplitude, while the resulting peak height in the Fourier 
spectrum will be referred to as Fourier amplitude. For each simulation, the mean S/N ratio of the 
six highest peaks was calculated (i.e., we assumed 6 out of 11 frequencies to be a successful 
detection of p-mode pulsation). The results of the simulations are shown in Figure \ref{fig:fig02}.

Not surprisingly, the simulations show that oscillations with larger amplitudes and longer 
mode lifetimes would be easier to detect. For example, an RMS amplitude of 16\,ppm of the oscillations 
would remain undetected if the mode lifetimes are very short, even though Fourier domain noise would 
in principle allow a $4\,\sigma$ detection. Only for lifetimes getting closer to the length of 
the dataset are the expected S/N ratios as known for a coherent signal fulfilled (e.g., a 
coherent sinusoidal variation with a RMS amplitude of 15\,ppm produces a peak amplitude of 
about 20\,ppm, corresponding to a S/N\,$\sim$\,5 in the 85\,Peg data). The positive influence of 
longer lifetimes, however, decreases rapidly as the lifetime approaches the length of the dataset, 
reaching a detection level of about 10\,ppm at $\tau>10$\,d. Note that in Equation 
\ref{equ:amp}, $H$ is only representative of the power excess height for large values of 
$T/\tau$, i.e. when the Lorentzian profile is well resolved. This explains why a signal 
formally described by, e.g., $H=18$\,ppm can remain undetected for long lifetimes 
(i.e. small $T/\tau$) in the MOST photometry simulations. With this in mind, 
the following conclusions on the oscillation properties of 85\,Peg can be drawn: 
1) Independent of the amplitude or the mode lifetime of the oscillations, the height of the Lorentzian 
profiles must be lower than about 20\,ppm to remain undetected in the MOST data; 2) If the mode 
lifetimes in 85\,Peg are very short ($\sim$ days), the RMS amplitude of the signal must lie below 
a level of about 12--18\,ppm. If the lifetimes are long ($\sim$ weeks), this limit lies at 
approximately 10\,ppm.

In order to predict upper limits for the height of the Lorentzian profiles for 85\,Peg\,A 
which are independent of the observation timebase, we converted the values derived from the simulations 
into power density. 
The top panel of Figure \ref{fig:fig04} shows, for each mode lifetime bin of the simulations, 
data points with S/N=3.5--4.5 and the corresponding power law fit in power density.
Combined with the Fourier amplitude limit of $\sim 16$\,ppm set by MOST, this provides firm limits on the oscillation 
properties of 85\,Peg\,A and could be used as a prior information to plan and analyze further observations 
with the goal of detecting solar-like oscillations in this star.

\subsection{Comparison with theoretical values}
In order to compare our upper limits on the oscillation amplitudes with predictions, 
we follow the method by \citet{michel_abol} to calculate bolometric amplitudes, which are 
independent of the instrument being used. Using the transmission curve of the MOST filter and 
an effective temperature of 5600\,K for 85\,Peg\,A, we calculate a response function for radial 
modes of $R_{0}=4.787$. Following the convention by \citet{michel_abol}, we then convert our 
upper RMS amplitude limits as derived in the previous section into bolometric amplitudes per radial 
mode as

\begin{equation}
A_{\rm{bol},l=0} = \frac{4}{R_{0}} \left(\frac{\delta I}{\overline{I}} \right ) (t) \,.
\end{equation}

\noindent
where the last term denotes the observed RMS intensity 
fluctuations, which we identify as our definition of the RMS amplitude.

Theoretical values calibrated to observations indicate that bolometric photometric 
amplitudes scale as 
$(L/M)^{\alpha} T_{\rm eff}^{-1}$ \citep{kjeldsen}, with numerical simulations suggesting a 
value of $\alpha\sim 0.7$ \citep{samadi}.
Using the solar reference value of $2.53\pm0.11$\,ppm for $A_{\rm{bol},l=0}$ 
by \citet{michel_abol}, the mass, luminosity and effective temperature (with a more 
conservative error of 100\,K) of \makebox{85\,Peg\,A} yield an expected theoretical amplitude of 
2.2$\pm$0.2\,ppm. As an estimate for the expected mode lifetime we use the recently suggested scaling 
relation by \citet{chaplin2009} proposing $\tau \propto T_{\rm eff}^{-4}$, and with a solar 
reference value of $\tau_{\sun}=3.2\pm 0.2$\,d \citep{chaplin2005} arrive at 
$\tau=3.6\pm 0.3$\,d for the expected mode lifetime of \makebox{85\,Peg\,A}.

\begin{figure}
\resizebox{\hsize}{!}{\includegraphics{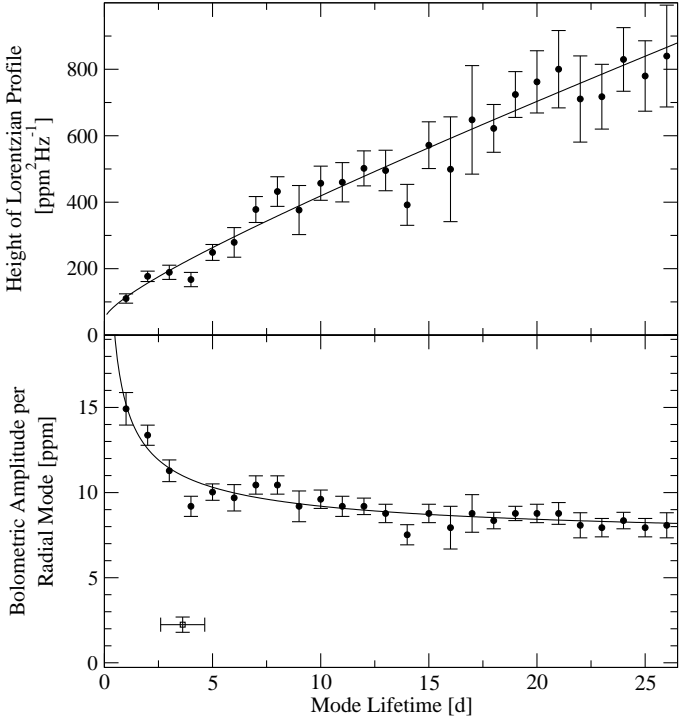}} 
\caption{Upper limits on oscillation properties of 85\,Peg\,A as a 
function of mode lifetime, based on simulations using MOST photometry. Top panel: Upper limit on the 
height of the Lorentzian profile in power 
density. Bottom panel: Upper limit on the bolometric amplitude per radial mode. The open 
square shows the expected value based on theoretical scaling relations with 3\,$\sigma$ error 
bars.}
\label{fig:fig03}
\end{figure}

The bottom panel of Figure \ref{fig:fig03} shows the upper limits for 85\,Peg\,A as derived from the 
simulations converted to bolometric amplitudes per radial mode, compared 
to the theoretical values. Clearly, the low value for the theoretical amplitude would in any case remain 
undetected in the MOST 85\,Peg data. 
This is also true for the mean velocity amplitude of 9\,cm\,s$^{-1}$ estimated by Houdek \citep[cf.][]{dantona} 
for a model of \makebox{85\,Peg\,A}, which, using the relation by \citet{kjeldsen}, would 
translate in a bolometric luminosity amplitude of $\sim$1\,ppm. Considering these estimated 
values, we are not able to draw any conclusions 
about the mode lifetimes for this star and can conclude that our results as upper limits are in agreement 
with predictions.

\subsection{Solar-type oscillations or wishful thinking?}

In a further investigation of the 2005 data, a relaxed spectral significance limit of 3 
(corresponding to a S/N of about 3) was used to search for p-modes in the 85\,Peg data. 
In this case, {\sc{SigSpec}} identified several peaks which were mostly located in two 
distinct frequency regions, one reaching from \makebox{1--3\,mHz}, the other from about \makebox{5--7\,mHz}. 
Eliminating all signals which are due to known instrumental artifacts and considering only values in the 
frequency range where p-modes are predicted, these frequency candidates were fitted to a dense 
grid of 300\,000 stellar models \citep{guenther02}. The quality of the fit was evaluated by a 
$\chi^{2}$ test, with a value of ${\chi}^2\leq 1$ meaning that, on average, the model frequencies 
that are used in the fit are within the uncertainties of the observed frequencies (see Guenther \& Brown 2004 
for details). This attempt resulted in a model fitting to about 65\% of the observed frequencies with a $\chi^{2}\leq 1$.  
This model is in agreement with the previously estimated position of \makebox{85\,Peg\,A} in the HR-Diagram, 
with a large frequency separation of 172\,$\mu$Hz.

\begin{figure}
\resizebox{\hsize}{!}{\includegraphics{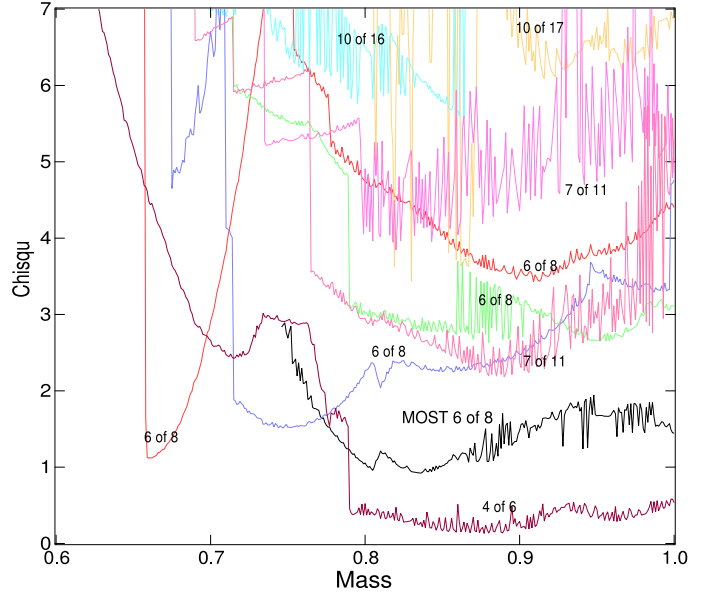}} 
\caption{Model fitting results as a function of mass using frequencies extracted from randomly generated datasets 
with a limit of S/N $\geq$ 3.  Line labels indicate the number of frequencies out of the total number of 
detected frequencies in a given dataset used for the fit. The fitting result using the real MOST dataset is also shown.}
\label{fig:fig04}
\end{figure}

To test the credibility of this model fit, we generated 8 simulated datasets, 4 using gaussian distributed 
random numbers, the other half by randomly shuffling the 2005 light curve. In both cases the same sampling, 
frequency analysis and model fitting procedure as for the real light curve was applied. The results of this 
test are shown in Figure \ref{fig:fig04}, with the fit quality $\chi^{2}$ given as a function of mass. 
As the Figure shows, there is indeed one random dataset where frequencies could be extracted that also fit 
with a ${\chi}^2\leq 1$, indicating a similar fit quality as the best model derived for the 
real data.

The result of this test underlines that when searching for low-amplitude p-mode oscillations, it is 
extremely important to discern real evidence from wishful thinking. In a recent analysis of MOST data 
on Procyon, \citet{guenther_procyon} have demonstrated the risks when searching for comb-like 
structures in power spectra of sun-like stars using a priori assumptions. The test presented here 
is yet another example for this, showing that a S/N limit of 3 in amplitude (hence S/N $\geq$ 9 in 
power, which is more widely used) is not sufficient to 
distinguish real signal from peaks generated from random noise, and can lead to misleading results.

\section{Long periodic variability}
\label{sec:long}

Although designed as a non-differential photometer, the MOST space telescope has proven to be 
remarkably stable over long timescales, making it also suited to search for long-periodic 
variability. Instrumental trends, if present, can in many cases be identified, reconstructed and 
corrected by analyzing the satellites telemetry data such as board or preamplifier temperature. 

\begin{figure}[h!]
\resizebox{\hsize}{!}{\includegraphics{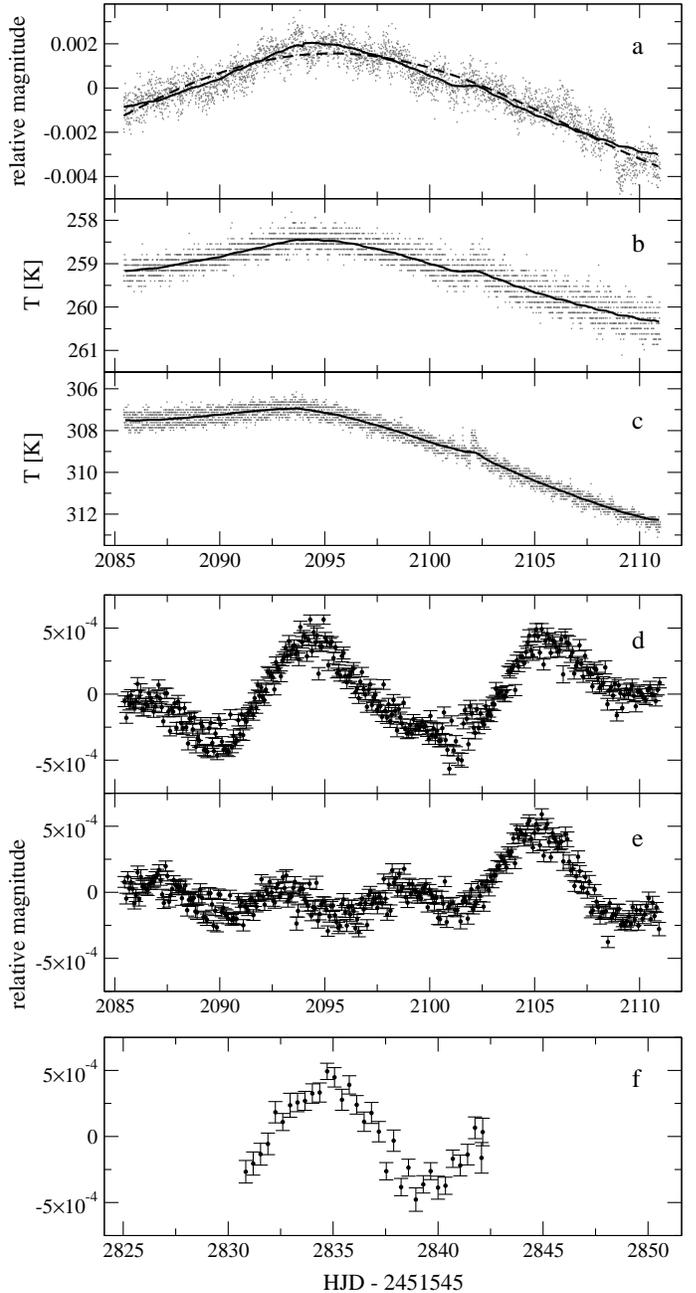}} 
%\resizebox{10.5cm}{!}{\includegraphics{fig05.eps}} 	
 \caption{Panel (a): Original 2005 85\,Peg light curve (every tenth data point is shown) 
together with different fits used to correct the instrumental trend (see text). Panel (b): 
Satellite preamplifier temperature curve (every tenth data point is shown). The solid line shows 
a smoothed version. Panel (c): Same as panel (b) but for satellite board temperature. 
Panel (d): Binned and low-pass filtered 2005 85\,Peg light curve derived by subtracting a polynomial 
fit (dashed line in panel (a)). 
Panel (e): Same as in panel (d), but derived by decorrelating variations found in the satellite 
telemetry data (solid line in Figure panel (a)). 
Panel (f): Corrected 2007 85\,Peg light curve. Note that the y-axis scale for panels (d), (e) and 
(f) is the same, but differs from the scale in panel (a).}
\label{fig:fig05}
\end{figure}

The 2005 light curve of 85\,Peg shows a dominant long-term instrumental trend which is confirmed 
by inspection of both the board and preamplifier temperature data. All three datasets are displayed 
in panels (a), (b) and (c) of Figure \ref{fig:fig05}. 
In a first attempt to correct the trend seen in panel (a), we fitted a polynomial function to the data 
(dashed line). 
The residuals of this fit, low-pass filtered to remove residual variability with periods of one day 
and averaged into bins of 
length 101\,min (about one MOST orbital period), show two cycles of a variability with a period of 
about 11\,d and an 
amplitude of 0.3\,mmag (panel (d) of Figure 
\ref{fig:fig05}). The 2007 data, which showed a less complex instrumental trend corrected by a simple 
linear fit, seems to confirm this variability detected in 2005, showing one full cycle of a light 
variation with a comparable period and amplitude (\makebox{Figure \ref{fig:fig05}}, panel (f)). Combining the 
2005 and 2007 data, a Fourier analysis yields a best fit period of 11.6\,d for the variation. 

In a later stage of the analysis, an attempt was made to correct the 2005 data trend by directly using 
and fitting the variations seen in the satellite telemetry data\footnote{Note that the brightest 
star simultaneously observed in the 2005 run is $\sim$\,3\,mag fainter and hence no comparison stars 
can be used for this correction.}. For this purpose, the telemetry data were binned to 
two orbital period means and heavily smoothed, so only long-periodic features remain present. These 
smoothed versions (shown as solid lines in panels (b) and (c) of Figure  \ref{fig:fig05}) were then 
correlated to the light curve in two segments of the run, and using the 
comparison set showing the highest correlation coefficient (in both cases the preamplifier 
temperature, see panel (b) of Figure \ref{fig:fig05}) all 
correlations were removed using linear regressions. The resulting fit which was subtracted is 
shown as a solid line in panel (a) of Figure \ref{fig:fig05}. The resulting light curve (panel (e) 
in Figure \ref{fig:fig05}) shows a much more complex behaviour, with the first cycle of the 11\,d 
variability almost entirely removed.

The considerable discrepancy between the results of the two background removal methods is obviously 
alarming. However, as the variability of the first reduction method was confirmed in the 2007 dataset, 
we feel that 
we have observed the $\sim 11$\,d periodicity in both epochs of data. We stress that without 
follow-up observations to confirm the observed variability we cannot rule out the possibility that 
the correlation between the 2005 and 2007 epochs of data is purely coincidental, and that the 
variability changed considerably between 2005 and 2007 as illustrated with our second reduction 
method. It must be noted that the usage of the decorrelation technique in this case assumes a 
linear relationship of temperature and intensity variability, which might be
an oversimplification of the problem. Indeed recent results on the low frequency variability of the 
MOST data on Procyon has shown evidence that telemetry data decorrelation
seems to overcorrect existing intrinsic variability (Huber et al., in preparation).
Thus we feel that the most likely possibility is that we have observed a common $\sim 11$\,d 
variability in our two epochs of data.  In the following sections we present attempts to interpret 
the variability assuming that the 2005 light curve in panel (d) of Figure 5 is indeed intrinsic.

The variability could originate in either of the two 
stars in the binary, since the MOST photometry does not resolve the individual components.
If it is due to star A, then the amplitude of the variability is 0.3\,mmag. If it is due to star B, 
then the amplitude we observe has been diluted significantly by the light of star A. The intrinsic 
amplitude would be approximately 5\,mmag, based on the visual magnitude difference of 85\,Peg\,A and 
85\,Peg\,B.

We have considered two possible explanations of the periodicity: (1) light effects due to the 
possible binarity of component B, and (2) 
rotational modulation of the light of either star A or B due to surface spots.

\subsection{Binarity of 85\,Peg\,B}

\makebox{85\,Peg\,B} has long been suspected to be a multiple system due to the discrepancy between the 
dynamically determined masses and the magnitude difference between stars A and B. The shape 
of the light curves in panels (d) and (f) of Figure \ref{fig:fig05} is not consistent with a simple detached eclipsing 
binary, but it could be due to tidal distortions or backheating in a slightly eccentric binary system 
with a period of 11.6\,d. Indeed, \citet{griffin} points out that the period of a putative 85 Peg Bb 
of less than a few weeks would be possible.

Based on the location of \makebox{85\,Peg\,B} in the HR diagram and by applying a mass-luminosity relation, 
\citet{dantona} suggested the luminosities of a possible secondary system to be $L_{Ba}\sim 0.064\,L_{\sun}$ 
and $L_{Bb}\sim 0.0085\,L_{\sun}$ (with the luminosity of the primary $L_{A}\sim 0.617\,L_{\sun}$). 
The masses of the components of the possible 85 Peg Bab binary are estimated to be $M_{Ba}\sim 0.5\,M_{\odot}$ 
and  $M_{Bb}\sim 0.2\,M_{\odot}$ \citep{bach,dantona}.
Therefore, an orbital period $P=11.6$\,d means a semi-major axis $a\sim 0.09$\,AU (about 
20\,$R_{\odot}$). If the stars in \makebox{85\,Peg\,B} are on the main sequence, then this separation would
be about $20 \times$ the expected diameter of star Ba. 

The amplitude of the light variation is about 5\,mmag (allowing for dilution by 
star A), which means the \makebox{\makebox{85\,Peg\,B}ab} binary must produce a peak-to-peak variation of about 1\%.  
If the light variation is primarily a geometrical effect of tidal distortion, the light curve would be a double 
wave (with two maxima and two minima) during each binary orbit. The latter would mean the 
binary period would be $2\times 11.6=23.2$\,d.  However, for such a period, the 
semi-major axis of the orbit would be $a\sim 0.14$\,AU, or about $30 \times$ the diameter of
star Ba. At this separation, the geometrical distortion of star Ba by tides would be far below 
1\% and could not account for the full amplitude of the observed variation. 

The bolometric flux on the surface of star Ba due to star Bb ($L_{Bb}\sim 0.0085\,L_{\sun}$ at 
a distance of about $0.1$\,AU) would be $\sim1.5\times 10^4$\,W/m$^2$. The surface flux of 
star Ba ($L_{Ba}\sim 0.064\,L_{\sun}$; $R\sim 0.6\,R_{\sun}$) would be roughly $1.4\times 
10^8$\,W/m$^2$. Therefore the flux ratio due to backheating by a putative star Bb is only about $10^{-4}$, 
and the expected peak-to-peak amplitude would be in the order of 120\,ppm. This is two orders of magnitude 
smaller than the observed peak variation in brightness, if that modulation originates in component B 
of the 85 Peg system.

Is it possible that the brightness ratio of stars A and B is not $\sim$16 due to the custom 
design of the MOST bandpass, and that variability due to star B would not be diluted as much as we have assumed?
We tested this possibility by generating stellar models across a range of masses, for ages
of 8\,Gyr, and determining their flux ratios measured through the MOST filter (Table \ref{table:fluxratios}).

\begin{table}  
\centering  
\caption{Model parameters for several possible configurations of the 85\,Peg system
and flux ratios as measured through the MOST custom filter. The deviations in temperature to the observed value 
given in \citet{dantona} are given in terms of the standard deviation $\sigma$. 
$A_{\mathrm{obs}}$ indicates the observed amplitude scaled with the corresponding flux ratio. The 
expected amplitude due to backheating of star Bb on star Ba is about 0.1--0.2\,mmag.} 
\begin{tabular}{c c c c c c c}        
\hline
\hline
Star 	&  $M$          	&  $R$        		&  $T_{\rm eff}$  	&	$\Delta T$ 		&	Flux ratio 	& $A_{\mathrm{obs}}$ 		\\
       	& ($M_{\sun}$)		& ($R_{\sun}$) 		&  (K)            	&   ($\sigma$) 		& A/B 	& (mmag)					\\														\\
\hline       
 A    	&  0.90            	&  0.92        		&   5460    		&	3				& 14.1        	& 4.5          			\\
 Ba  	&  0.55            	&  0.48        		&   4200         	&   0            	&				&						\\ 
 Bb   	& -           		&  -  		   		&   -   	    	&   -        		&				&						\\
\hline
 A    	&  0.77            	&  0.74        		&   4845        	&	15				& 1.5 			& 0.48					\\   
 Ba  	&  0.72            	&  0.68        		&   4605          	&	2				&               &						\\ 
 Bb   	& -            		&  -        		&   -        		&	-				&               & 						\\
\hline
 A    	&  0.90           	&  0.92          	&   5460          	& 	3				& 15.2			& 4.9					\\
 Ba   	&  0.52          	&  0.46          	&   4160          	&   0.2				&				&			         \\
 Bb   	&  0.20          	&  0.21          	&   3650           	&   -				&				&			            \\
\hline
 A    	&  0.77            	&  0.74        		&   4845          	&  	15				& 5.4 			& 1.7  		 	       \\   
 Ba  	&  0.52            	&  0.46        		&   4160           	&   0.2   			&				&						\\ 
 Bb   	& 0.20            	&  0.21        		&   3650           	&   -				&				&						    \\
\hline               
\end{tabular}  
\label{table:fluxratios} 
\end{table}

To change the flux ratio measured by MOST significantly, the temperature of star A would have to be considerably 
cooler than measured for this relatively bright star. Even if the the ratio is lowered by a factor 
of 3 (last row of Table \ref{table:fluxratios}), the observed amplitude would still be (in the most 
optimistic case) by about a factor of 9 too low. Therefore, it is unlikely that illumination of a 
component Ba by a putative star Bb can easily account for the amplitude of the observed variation.

\subsection{Rotational modulation}

\subsubsection{Determination of rotational velocity}
\label{sec:vsini}
In order to interpret the variation in terms of rotational modulation, a good knowledge of the 
projected rotational velocity $v\sin i$ is essential. Searching the literature we found 
measurements in the catalogues of \citet{bernacca} and 
\citet{usuegi} giving $v\sin i$ values of $3\pm 3$ and $<6$\,km\,s$^{-1}$, respectively, for \makebox{85\,Peg\,A}.  
More recently, \citet{wolff} measured $v$\,sin$i\simeq 4$\,km\,s$^{-1}$ and \citet{hale} gives a value of 
$1.8\pm 0.6$\,km\,s$^{-1}$. The discrepancy between all these measurements is evident, and hence 
$v\sin i$ of \makebox{85\,Peg\,A} remains a rather poorly determined parameter. Note that due to the close 
separation, no measurements for \makebox{85\,Peg\,B} are available.

With the goal to arrive at an independent $v\sin i$ measurement including statistically meaningful 
uncertainty estimates, we extracted high-resolution ($\rm{R}\sim60000$) spectra taken with the UVES spectrograph 
from the ESO archive which were previously used to study star formation in the galaxy through 
abundance ratios (project number 076.B-0133). 

\begin{figure}[h!]
\resizebox{\hsize}{!}{\includegraphics{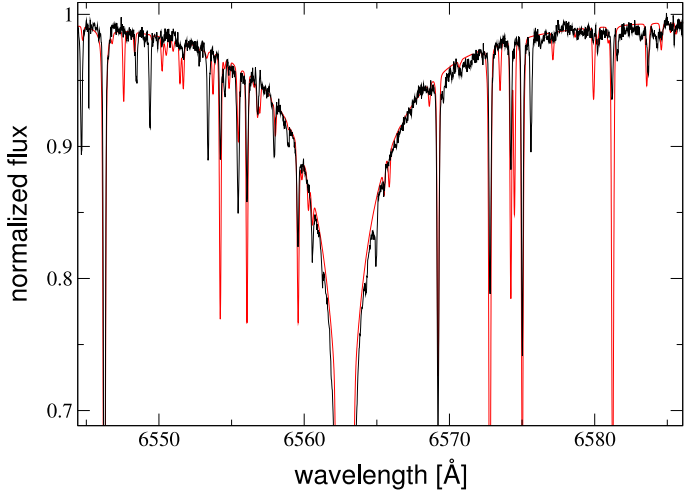}} 
\caption{Observed UVES spectrum (thick black line) and fitted synthetic spectrum with $\log g=4.5$ and 
$T_{\rm eff}=5500$\,K (thin red line).}
\label{fig:fig06}
\end{figure}

After the pipeline reduction and continuum normalization, synthetic spectra were computed 
using the program synth3 \citep{synth3} in conjunction with VALD \citep{vald1,vald2,vald3}. The fitting of the syntheses to the 
observed H$\alpha$ line, which can be seen in Figure \ref{fig:fig06}, resulted in $\log g=4.5$ and an effective 
temperature $T_{\rm eff}=5500$\,K. 

\begin{figure}[h!]
\resizebox{\hsize}{!}{\includegraphics{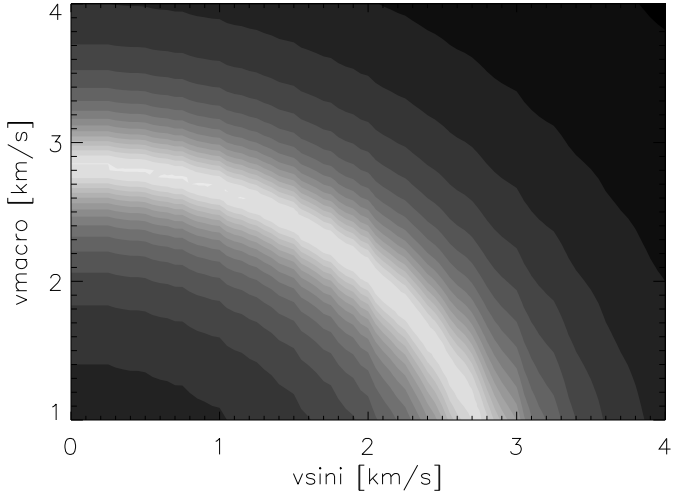}} 
\caption{The goodness-of-fit of observed to synthetic spectrum of one Fe line as a function of 
$v\sin i$ and $v_{\rm mac}$. Bright areas indicate a good fit.}
\label{fig:fig07}
\end{figure}
%michis

Based on this effective temperature, MARCS model atmospheres \citep{marcs} were chosen for the 
subsequent analysis. 
After the equivalent widths were determined using a modified version \citep{vadim} of the WIDTH9 code 
\citep{kurucz1993a}, a microturbulence velocity ($v_{\rm mic}$) of 1.3\,km\,s$^{-1}$ was determined 
by eliminating any abundance-equivalent width correlation. Specific Fe lines with highly accurate 
$\log g_{\rm f}$ (T. Ryabchikova, private communication) were used for the $v\sin i$ determination. An Fe abundance of 
$-\,0.35\,\pm\,0.05\,dex$ relative to the solar value \citep{asplund} and a radial velocity of 34.6\,km\,s$^{-1}$ 
were adopted to enable a good comparison of observation and synthesis. Since the effects of macroturbulence 
velocity ($v_{\rm mac}$) and $v\sin i$ on the goodness-of-fit in the spectrum are indistinguishable 
to the naked eye (see Figure \ref{fig:fig07}), we pursued to approximate the marginal distribution of $v\sin i$. 

For this purpose, we calculated a 2D-grid of synthetic spectra, using the parameter values mentioned 
above, but with $v\sin i$ ranging from 0 to 4\,km\,s$^{-1}$ and $v_{\rm mac}$ ranging from 1 to 
4\,km\,s$^{-1}$ in steps of 0.05. Subsequently, we compared this grid of models to the four most 
clearly defined iron lines in the observed spectrum. However, each of these lines was treated as an 
independent observation to overall create four probability distributions. The error bars of each observed 
data point, as delivered by the reduction pipeline, were interpreted as $1\sigma$-confidence limits 
and the probability for each model value was calculated according to 

\begin{equation}
p(d_{\rm O, i}) = \exp\left({-\frac{(d_{\rm O, i} - d_{\rm M, i})^2}{2{\sigma_{\rm i}}^2}}\right),
\label{equ:prob}
\end{equation}
\noindent
were $d_{\rm O, i}$ is the i-th observed data point, $d_{\rm M, i}$ is the corresponding model value, 
and $\sigma_{\rm i}$ the aforementioned $1\sigma$-uncertainty. The probability $p_{\rm M}$ for each 
model then is the product of the probabilities for all data points. 

\begin{figure}[t!]
\resizebox{\hsize}{!}{\includegraphics{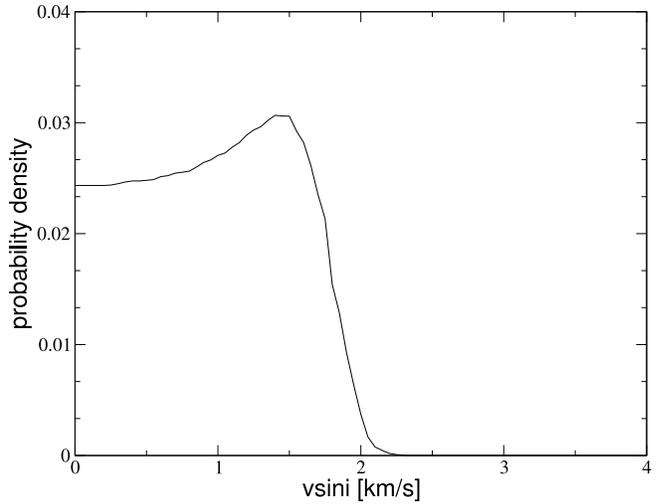}} 
\caption{The marginal distribution of the $v\sin i$ parameter derived as described in $\S$\ref{sec:vsini}. 
There is no evidence for a $v\sin i$ above 2.5\,km\,s$^{-1}$ according to the UVES data we used for our analysis.}
\label{fig:fig08}
\end{figure}

We calculated the marginal distributions of $v\sin i$ for each line by integrating out the $v_{\rm mac}$ 
parameter. Since we used a discrete parameter space in our 2D-grid, we summed up the model probabilities 
over the whole parameter range of $v_{\rm mac}$ for each specific value of $v\sin i$, and normalized the 
distribution so that the sum of all resulting probabilities is unity. Finally, we combined the marginal 
distributions of all Fe lines by calculating the joint probabilities for all models and, again, 
normalized the resulting distribution. The final distribution is shown in Figure \ref{fig:fig08}. 
It was subsequently used as a prior\footnote{In Bayesian probability analysis, the prior is a 
probability distribution conveying already existing knowledge about a quantity. The information 
contained in the prior is used to constrain and assess the final probability of the outcome (see 
Bayesian inference textbooks for details).} for the $v\sin i$ parameter in the spot model analysis described in the 
next section. Judging from the distribution alone, we obtain $v\sin i=1.40^{+0.15}_{-1.10}$ (with the 
uncertainties roughly corresponding to a 1$\sigma$ confidence limit).

\subsubsection{Spot model fitting}
To test the possibility of rotational modulation, we fit the 2005  and 2007 light curves 
(panels (d) and (f) of Figure \ref{fig:fig05}) with simple spot 
models. We employ the StarSpotz functionality discussed in \citet{croll_starspotz} and \citet{croll06a}, 
which adopts circular spots 
and applies analytic models described by \citet{budding}. StarSpotz has been used to 
measure the differential rotation profile of another MOST target, the young active solar-type star 
${\kappa}^1$\,Ceti \citep{walker07} and it was able to account for the photometric variations
detected by MOST and constrain the differential rotation coefficient  
of the exoplanet-hosting star $\epsilon$\,Eri \citep{croll06b}.  The reader is 
reminded at this point that in the following modeling efforts we explicitly assume that the variation seen in 
panel (d) of Figure \ref{fig:fig05} is intrinsic, and disregard evidence for more complex variability 
as seen in panel (e) of Figure \ref{fig:fig05}.

%\begin{figure*}
%\begin{center}
%\includegraphics[width=6cm]{phase_A_2005_final_I_hope.eps}
%\includegraphics[width=6cm]{phase_A_2007_final_I_hope.eps}
%\end{center}
%\begin{center}
%\includegraphics[width=6cm, angle=270]{model_resid_2005_A.eps}
%\includegraphics[width=6cm, angle=270]{model_resid_2007_A.eps}
%\end{center}
%\caption{Spot models of the periodic light variation observed by MOST in the 85\,Peg system if caused by
%85 Peg A. Top: The best fitting one-spot configurations for 85\,Peg\,A with the spot directed towards the viewer for
%2005 (left) and 2007 (right); the small dot is the modeled spot.
%Bottom left: The top panel shows the observed (binned) 2005 light curve and the model fit for 85\,Peg\,A, while the 
%bottom panel displays the residuals to the fit. Bottom right: Same as the middle panel except for the 2007 light
%curve.}
%\label{fig:spotmodelA}
%\end{figure*}

\begin{figure*}
\begin{center}
\includegraphics[width=6cm]{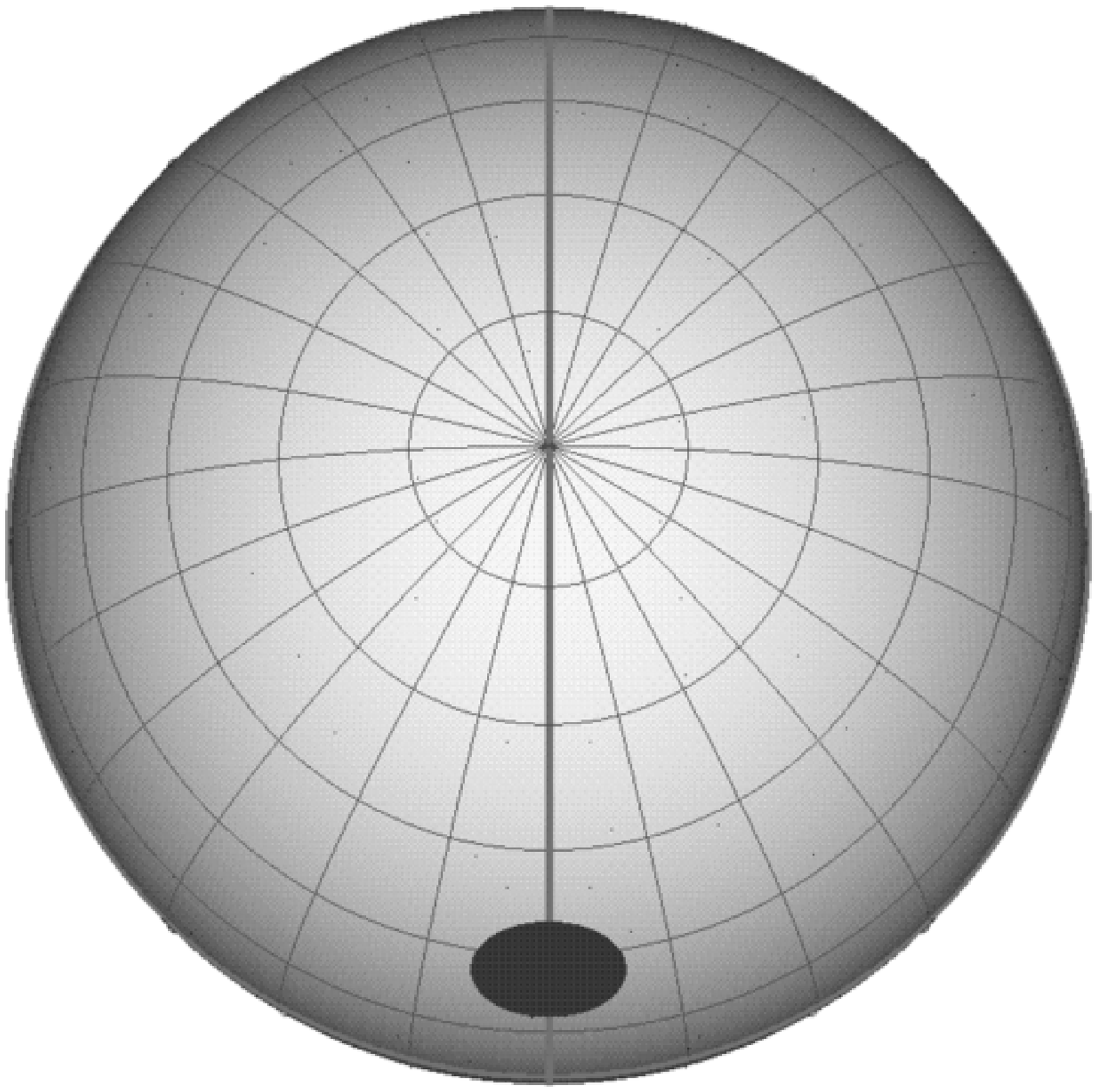}
\includegraphics[width=6cm]{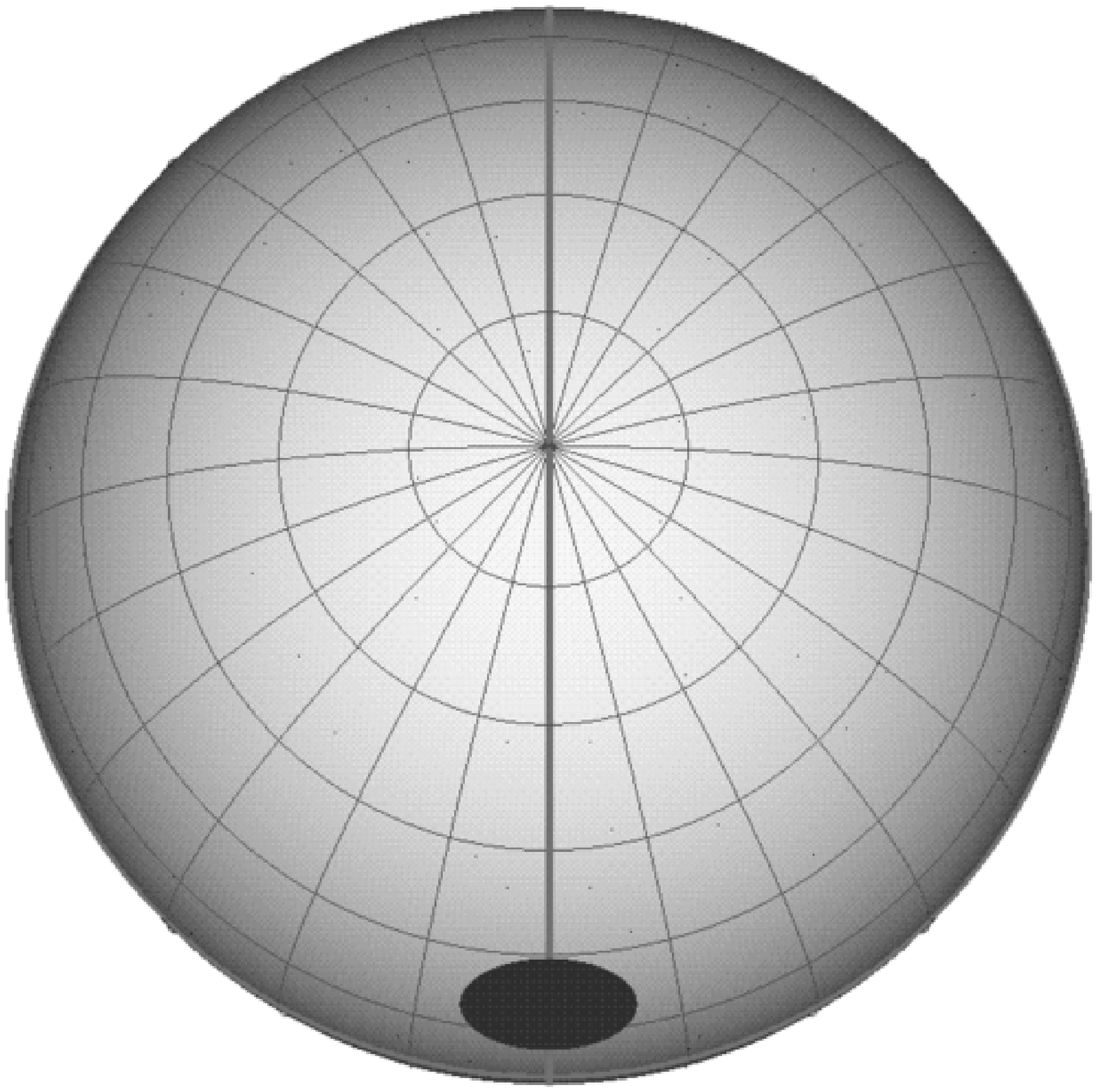}
\end{center}
\begin{center}
\includegraphics[width=6cm, angle=270]{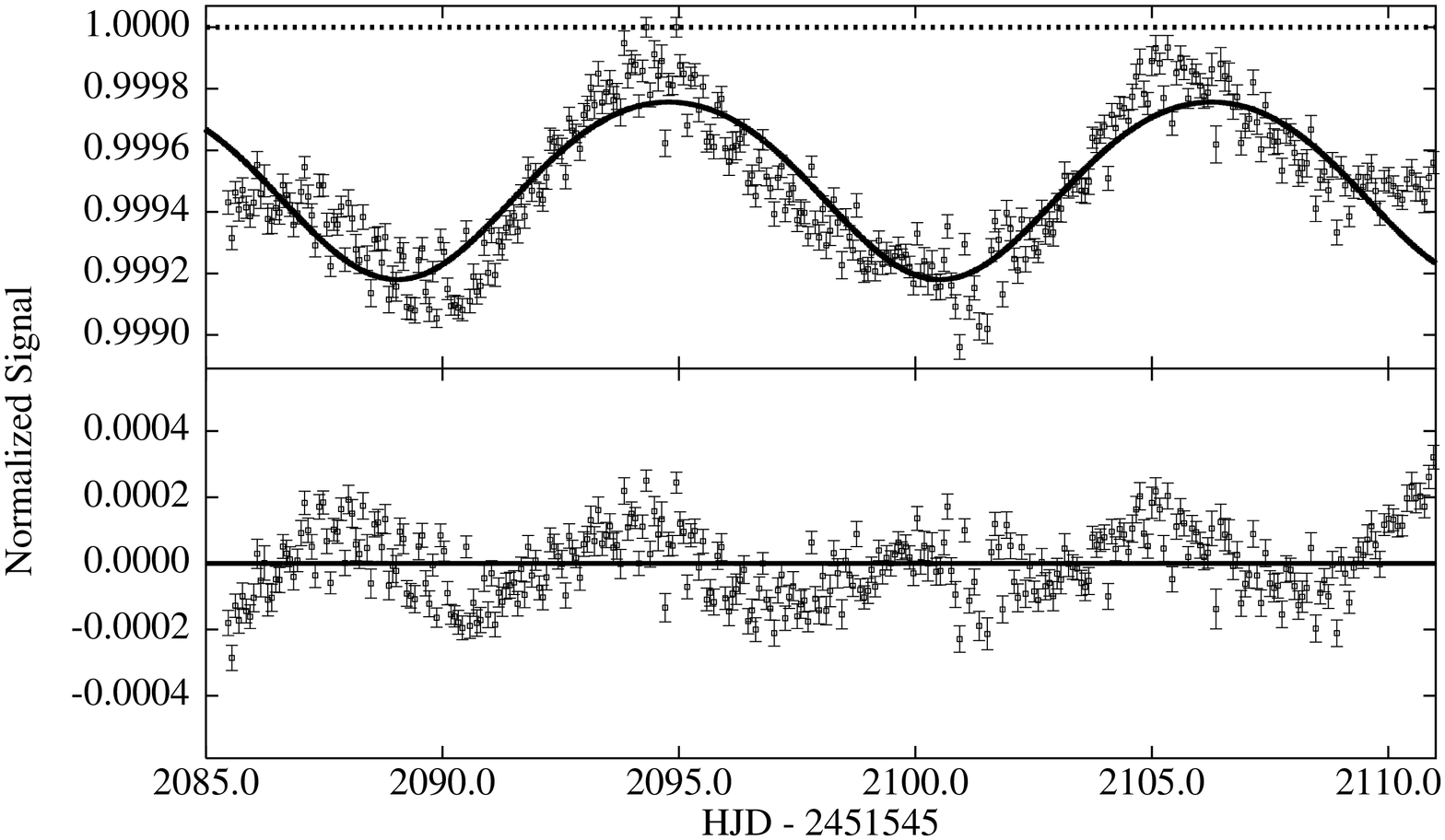}
\hspace{0.1cm}
\includegraphics[width=6cm, angle=270]{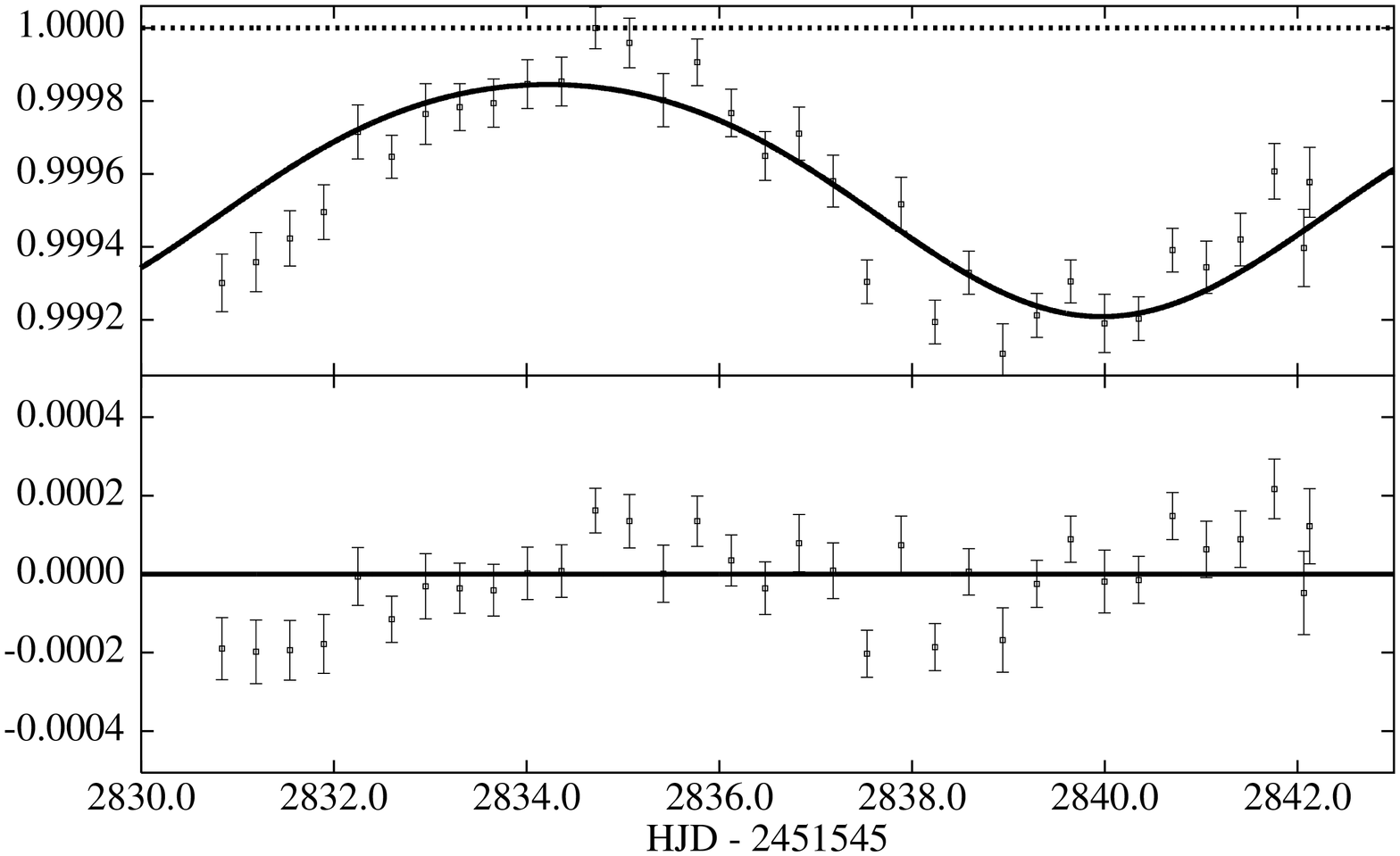}
\end{center}
\caption{Spot models of the periodic light variation observed by MOST in the 85\,Peg system if caused by
85 Peg B. Top: The best fitting one-spot configurations for 85\,Peg\,B with the spot facing the viewer for
2005 (left) and 2007 (right)
Bottom left: The top panel shows the observed (binned) 2005 light curve and the model fit for 85\,Peg\,B, while the 
bottom panel displays the residuals to the fit. Bottom right: Same as the middle panel except for the 2007 light
curve.}
\label{fig:fig09-12}
\end{figure*}

A separate spot model was generated for components A and B of the \makebox{85\,Peg} system. Linear limb darkening 
coefficients for both stars were estimated using the closest grid points to the physical parameters 
from \citet{dantona} in the catalog of \citet{claret}; these were $u\sim 0.86$ for \makebox{85\,Peg\,A} and 
$u\sim 0.80$ for \makebox{85\,Peg\,B}. Note that considering the uncertainty of the 
observed variation and the fact that we are only intending to provide a first 
approximation to model the observed variation, we refrain from using a more complex limb-darkening 
model in this application.
We used a flux contrast between the spotted and unspotted photosphere of $\kappa_{w}=0.30$ 
for \makebox{85\,Peg\,A} and $\kappa_{w}=0.35$ for \makebox{85\,Peg\,B}. 
We are unable to independently determine $\kappa_{w}$, and small variations in $\kappa_{w}$ will only
serve to decrease or increase the size of the modeled spots.
The value for \makebox{85\,Peg\,A} is appropriate for a G5IV star as it is comparable to that of 
the Sun and for the value chosen for $\kappa^{1}$\,Ceti \citep{walker07}. We assumed a similar value 
for the K6-8V star \makebox{85\,Peg\,B}. For the \makebox{85\,Peg\,B} model, the light curve was 
scaled to an amplitude of 5\,mmag to allow for the dilution by star A. 
For our \makebox{85\,Peg\,A} spot-model fit we place a priori constraint on the resulting $v\sin i$ 
value (thus a joint prior information on the rotation period and the rotational inclination angle of the star) 
based on the values calculated in $\S$\ref{sec:vsini}. For \makebox{85\,Peg\,B} we do not 
use such a prior constraint since the $v\sin i$ value of this target is unknown.

If the modulation is due to a single spot rotating on star A, the near-sinusoidal shape
and low amplitude (0.3\,mmag) of the light curves require this spot to be small (for the adopted flux 
contrast) and for it to be
visible for the duration of the light curve. Our determination of $v\sin i$ rules out edge-on and near edge-on 
inclinations for \makebox{85\,Peg\,A}; however, the uncertainty in $v\sin i$ is still significant and
thus our spot models of this star could vary from near pole-on inclination angles to much more moderate 
values ($i\sim 30^{\circ}$). Also, the observed modulation could be due to many spots or more 
complicated spot geometries. The spot model that follows should thus be in no way considered unique, 
but simply an example of a possible one-spot spot model that could be causing the observed modulation.

The best model for star A using the 2005 data features a spot of diameter $2^{\circ}$ at a latitude 
of about $50^{\circ}$, while the 2007 data features a spot of diameter $2^{\circ}$ at a latitude of 
about $46^{\circ}$ relative to the equator. Our spot model has a rotational inclination angle of $i\sim 21^{\circ}$. 
Note that since changing the inclination angle of \makebox{85\,Peg\,A}
does not significantly affect the $\chi^{2}$ of the spot model, our best fit model has an inclination angle
that results in a $v\sin i$ similar to the mean of the distribution determined by our analysis in 
$\S$\ref{sec:vsini}.
For the radius of $R_{\rm A}=0.82\pm 0.02\,R_{\odot}$ -- modeled by 
\citet{dantona} for the range of possible ages of \makebox{85\,Peg\,A} and close to the value of 
$0.846\,R_{\odot}$ estimated by \citet{fernandes} -- and a rotation period $P=11.6$\,d, 
the equatorial rotation velocity of the star would be 
$3.6$\,km\,s$^{-1}$. The spot model inclination results in a value of $v\sin i$ of about 1.3 km/s 
(compared to our $v\sin i\sim 1.40^{+0.15}_{-1.10}$ value determined in $\S$\ref{sec:vsini}).

For star B, the intrinsic amplitude of the light variation should be closer to 5\,mmag, requiring a 
larger spot if the contrast ratio between the spotted and unspotted photosphere is similar to that of other
K6-8V stars. The best-fit model is presented in Figure \ref{fig:fig09-12} (note that the best-fit 
model for star A is very similar, except with a larger 
inclination angle and smaller spot diameter).
In this case, the inclination of the star for the best fitted model is $i \simeq 11^{\circ}$. In 2005 
the spot in the best-fit model has
a diameter of $8^{\circ}$ at a latitude of $27^{\circ}$. In 2007 the spot has a diameter of $9^{\circ}$ at a latitude of 
$20^{\circ}$. The equatorial rotation velocity of star B would be about $4.7$ km/s and the 
resulting $v\sin i$ of the model is about $0.9$\,km\,s$^{-1}$. We could find no measurements of 
$v\sin i$ for \makebox{85\,Peg\,B} in the literature.

\section{Conclusions}
Nearly continuous photometry of the binary system 85\,Peg obtained with the MOST space telescope in 
2005 yields a null result for the detection of p-modes in the metal-poor subdwarf \makebox{85\,Peg\,A}. 
Spanning over 25 days and reaching down to a noise level of about 4\,ppm in the frequency region of interest, 
the high precision data present the most complete photometry of the system collected so far and 
enable us to set an upper limit for the detection of such oscillations. Based on simulations of 
stochastically excited and damped p-mode signal, we conclude that the RMS amplitude of oscillations 
in 85\,Peg\,A must lie below $\sim$\,12--$18$\,ppm if the mode lifetimes are short (i.e. on the order of 
days) and below $\sim$\,10\,ppm if the mode lifetimes are long (i.e. on the order of 
weeks). Based on these results, we provide upper limits for the Lorentzian profile height and 
the bolometric amplitude per radial mode as a function of mode lifetime and find these limits to be 
in agreement with theoretical scaling relations. Due to the fact that 
the expected theoretical amplitude is a factor 4--5 lower than the detection threshold, however, 
we are unable to comment further on the comparison of theoretical and observed amlitudes as 
recently done by \citet{michel_science} for CoRoT observations.

We detect light variations in the 85 Peg system with a period of about 11\,d and a peak amplitude 
of 0.3\,mmag in our 2005 epoch of data and confirm this variability in the 2007 epoch of data. However, 
as different methods to remove the instrumental background in the 2005 data yield vastly different 
results, we cannot exclude the possibility that we do not know the exact shape and period of the 2005 
variability. Nevertheless, as the 2005 and 2007 data yield similar results we believe it is likely 
that 85 Peg displayed consistent variability over these two epochs.

The observed peak amplitude of the periodicity can in any case not be explained by effects of a 
possible undetected binary \makebox{85\,Peg\,B}. Ignoring correlations with telemetry data which exist 
but cannot be interpreted beyond doubt, we fitted simple models 
to the residual data under the assumption that the variation is caused by star spots modulated with the 
rotation period of one of the components. To facilitate the search for such models we analyzed 
high-resolution UVES spectra of 85\,Peg found in the ESO archive and derived a new estimate of 
$v\sin i=1.40^{+0.15}_{-1.10}$ for \makebox{85\,Peg\,A}. While we are able to reproduce the light curves 
of both datasets well with single-spot models, it must be emphasized that the solutions presented here 
are by no means unique, and there is no evidence that rules out more complex models.

Assuming that our simple models are realistic, this would indicate a rotation period of 11.6\,d 
for either one of the components. \citet{noyes} derived a rotation period for \makebox{85\,Peg\,A} of 
23.8 days indirectly based on its chromospheric activity index. The evolutionary models of \citet{dantona} 
allow ages from 8--14\,Gyr, while those of \citet{bach} result in an age of $8.4\pm 0.5$\,Gyr for both 
stars in the 85 Peg system. Component B is expected to be coeval with star A and 
should share the same age. All of these age estimates suggest that the stars 
should have a slower rotation period than the observed variation shows. For example, the much younger 
active G5 dwarfs ${\kappa}^1$\,Ceti \citep{walker07} and HD\,189733 
\citep{henrywinn08} have rotation periods of 9 and 11.8 days, respectively - comparable to the 11.6 days 
rotation period suggested for 85 Peg despite the fact that the 85 Peg system is much older and should have a much 
longer rotation period.

The spectroscopic binary solution summarized by \citet{griffin} yields 
an orbit inclination of $|\,i\,|\sim 50^{\circ}$. While we do not explicitly assume a spin-orbit coupling in the 
system to constrain the inclination in the spot models described above, it is worth noting that 
for \makebox{85\,Peg\,A}, a rotation period of twice the observed period (23.2\,d) with such an inclination 
would result in a $v\sin i$ of 1.4\,km\,s$^{-1}$, which is also consistent with the values derived in 
$\S$\ref{sec:vsini} and might be a more realistic rotation period (suggesting that indeed a 
more complex spot model is required to explain the variation). Another, maybe more exotic 
explanation for the short period might be that the rotational modulation is produced in star B (the cooler and 
potentially more active component) and that its rotation is synchronized with the long suspected 
binary companion with an orbital period of 11.6\,d. 

Lastly, we strongly encourage follow-up observations to confirm the 11.6\,d observed 
variability.

\section*{Acknowledgments}
{We are grateful for discussions with Pierre Demarque. We are also thankful to Gerald Handler for 
critical discussions concerning the p-mode detection limit. MG, MO, TK, WWW and RK 
are supported by the Austrian Fonds zur F\"orderung der wissenschaftlichen Forschung, project number P17580-N02. 
The Austrian participation in the MOST project is funded by the Austrian Research Promotion Agency (FFG). 
BC, DBG, JMM, AFJM, and SR acknowledge funding from the Natural Sciences \& Engineering Research Council 
(NSERC) Canada.}

\bibliographystyle{aa}
\bibliography{references}

\end{document}